# Coexistence of static magnetism and superconductivity in SmFeAsO$_{1-x}$F$_x$ as revealed by muon spin rotation

A. J. Drew[1,2,*], Ch. Niedermayer[3], P. J. Baker[4], F. L. Pratt[5], S. J. Blundell[4], T. Lancaster[4], R. H. Liu[6], G. Wu[6], X. H. Chen[6], I. Watanabe[7], V. K. Malik[1], A. Dubroka[1], M. Rössle[1], K. W. Kim[1], C. Baines[8] and C. Bernhard[1,§]

[1] University of Fribourg, Department of Physics and Fribourg Centre for Nanomaterials, Chemin du Musée 3, CH-1700 Fribourg, Switzerland
[2] Queen Mary University of London, Department of Physics, Mile End Road, London E1 4NS, United Kingdom
[3] Laboratory for Neutron Scattering, Paul Scherrer Institut & ETH Zürich, CH-5232 Villigen, Switzerland
[4] Oxford University, Department of Physics, Clarendon Laboratory, Oxford OX1 3PU, United Kingdom
[5] ISIS Facility, Rutherford Appleton Laboratory, Chilton, Oxfordshire OX11 0QX, United Kingdom
[6] Hefei National Laboratory for Physical Sciences at Microscale and Department of Physics, University of Science and Technology of China, Hefei, Anhui 230026, China
[7] RIKEN-RAL, Nishina Centre, 2-1 Hirosawa, Wako, Saitama, 351-0198 Japan
[8] Laboratory for Muon Spin Spectroscopy, Paul Scherrer Institut, CH-5232 Villigen, Switzerland
[*] e-mail: A.J.Drew@qmul.ac.uk  [§] e-mail: christian.bernhard@unifr.ch

**The recent observation of superconductivity with critical temperatures (T$_c$) up to 55 K in the pnictide RFeAsO$_{1-x}$F$_x$, where R is a lanthanide marks the first discovery of a non copper-oxide based layered high T$_c$ superconductor (HTSC) [1 - 3]. It has raised the suspicion that these new materials share a similar pairing mechanism to the cuprate superconductors, since both families exhibit superconductivity following charge doping of a magnetic parent material. In this context, it is important to follow the evolution of the microscopic magnetic properties of the pnictides with doping and hence to determine whether magnetic correlations coexist with superconductivity. Here we present a muon spin rotation study on SmFeAsO$_{1-x}$F$_x$, with x=0 to 0.30 which shows that, as in the cuprates, static magnetism persists well into the superconducting regime. This analogy is quite surprising since the parent compounds of the two families have rather different magnetic ground states: itinerant spin density wave for the pnictides contrasted with the**

**Mott-Hubbard insulator in the cuprates. Our findings therefore suggest that the proximity to magnetic order and associated soft magnetic fluctuations, rather than strong electronic correlations in the vicinity of a Mott-Hubbard transition, may be the key ingredients of HTSC.**

Similar to the cuprates, the pnictide HTSC have a layered structure comprising alternating FeAs and LaO sheets, with the Fe arranged on a square lattice [1]. Theoretical calculations predict a quasi two-dimensional electronic structure, with LaO layers that mainly act as blocking layers and metallic FeAs layers that are responsible for superconductivity [4 - 6], although these are multiband superconductors with up to five FeAs-related bands crossing the Fermi-level [4 - 7]. Like the copper-oxide HTSCs, the superconducting state in the pnictides emerges upon charge doping a magnetic parent compound [8 - 10], with indications that the maximal $T_c$ occurs just as magnetism disappears [11 - 13]. The latter point may well be of great significance, since the parent compounds in the two families are very different. For the pnictides there are strong indications that they are itinerant systems with magnetism arising from a nesting-induced spin density wave (SDW). This is in contrast to the cuprates, where it is well established that the mother compounds are 'charge transfer insulators', where strongly repulsive electronic correlations yield an insulating and antiferromagnetic ground state despite a half-filled conduction band. It is therefore of great importance to obtain further insight into the differences and similiarities of the pnictide and cuprate HTSC. A particularly important question is how magnetism and superconductivity evolve upon electron doping. In this context, µSR is an ideal technique since it provides microscopic information corresponding to the bulk of a sample and there is a substantial body of µSR data that has been collected on the copper-oxide HTSC for comparison [14 - 16].

Here we present our zero-field muon spin rotation (ZF-µSR) measurements which detail how magnetism evolves with F-doping from the SDW state below $T_{mag} \approx 135$ K in the parent compound at x=0 towards the superconducting state at x≥0.1. In particular, we show that static magnetic correlations originating from the FeAs layers survive to a surprisingly high doping level well into the superconducting regime. Figures 1(a) to 1(e) display representative ZF-μSR spectra which show the time dependence of the muon spin polarisation, $P(t)/P(0)$, at temperatures below and above $T_{mag}$ in the order of increasing F content for $0 \leq x \leq 0.13$. The solid lines show fits to the experimental data with the function:

$$P(t) = f_1\left(\frac{2}{3}G_{osc} + \frac{1}{3}\right)\exp(-\lambda_1 t)^{\beta_1} + f_2 \exp(-\lambda_2 t)^{\beta_2} + f_{bg} \quad (1)$$

The first term with a fraction $f_1$ represents the static or quasi-static magnetic signal, the second term with a fraction $f_2$ the dynamic magnetic signal and the third term a fixed background component due to about 10% of the muons that miss the sample and stop in the sample holder or the cryostat walls. For x=0 the oscillatory part of the magnetic signal takes the form $G_{osc} = \cos(\omega_0 t)\exp(-\sigma^2 t^2)$, whereas $G_{osc} = J_0(\omega_0 t)\exp(-\sigma^2 t^2)$ for x>0. Here $\omega_0$ is a characteristic precession frequency, $\sigma$ is a Gaussian relaxation rate reflecting the frequency width and $J_0$ is a zeroth-order Bessel function. The parameters $\lambda_{1,2}$ and $\beta_{1,2}$ describe the generalised relaxation of the first two components due to magnetic fluctuations. The Bessel function used for x>0 reflects the intrinsic distribution of internal fields associated with an incommensurate spin density wave (SDW) [17]. The temperature and doping dependences of the fitted parameters are summarised in Figs. 2(a) – 2(c). They establish that bulk and static or quasi-static magnetism persists to a doping level of about x=0.13 and thus survives well into the superconducting regime.

In the first place, this raises the question whether the magnetism originates from electronic moments within the FeAs layers or is rather due to the ordering of the Sm moments. Our ZF-μSR data establish that all samples for x≤0.13 exhibit fairly high magnetic transition temperatures of $T_{mag} \geq 30$ K. This is a clear indication that the magnetic moments originate from the FeAs layers and are not associated with the Sm ions which order at much lower temperature [18]. The absence of a significant contribution of the Sm moments to the magnetic order at T≥10 K is also confirmed by our ZF-µSR data on the parent material SmFeAsO, whose precession frequency of 23.6 MHz (see Fig. 2c) is very similar to the frequency observed in LaOFeAs [10]. The static magnetism above 5 K is therefore due to magnetic moments that originate from the FeAs layers. As outlined in the supplementary material the ordering of the Sm moments below 5 K is evident in the ZF-µSR data, as well as in our specific heat data for the x=0.10 to 0.13 samples.

To obtain a quantitative analysis of the magnetic volume fractions we have performed additional weak transverse field (wTF-µSR) measurements. In a wTF, the magnetic fraction is not affected by the applied field *B* and only the non-magnetic terms in (1) precess in response to the field, giving the following polarisation function:

$$P(t) = f_1\left(\frac{2}{3}G_{osc} + \frac{1}{3}\right)\exp(-\lambda_1 t)^{\beta_1} + f_2 \exp(-\lambda_2 t)^{\beta_2} + f_{bg}\cos(\gamma_\mu Bt) \qquad (2)$$

Any non-magnetic fraction gives rise here to a weakly damped oscillatory signal whose amplitude can be readily determined. For all samples with x≤0.13 we find that only 10-15% of the muons experience a non-magnetic environment. These are mostly accounted for by the muons that stop outside the sample, either in the sample holder or in the walls and windows of the cryostat. The resulting doping dependence of the magnetic volume fraction is plotted in the inset of Figure 2(b). It highlights that for all samples with x≤0.13 the muons experience static magnetic fields in at least 90% of the sample volume. This does not necessarily imply that the static magnetic moments exist here in every unit cell. Nevertheless, since the stray fields from AF regions are known to decrease very rapidly, our data suggest that the spatial extent of such non-magnetic regions would be of the order of a few nanometers and thus of the SC coherence length as deduced from the upper critical field [22,23] and the gap magnitude [20].

Next we discuss the complementary issue of the SC volume fractions. Transverse field (TF-µSR) measurements of the vortex state, while they establish a nearly 100 % SC volume fraction at x=0.15 to 0.3 [19], cannot help at x≤0.13 since the relaxation behaviour here is dominated by the magnetism that is already present at at $T_c$. Nevertheless, Figures 3a and b show that the resistivity exhibits a fairly sharp drop towards zero and there is a sizeable diamagnetic signal in the magnetisation below $T_c$. In particular, in Figure 3c the gap-like suppression of the infrared optical conductivity, $\sigma_1$, is a clear signature of bulk superconductivity. This suppression is less pronounced at x=0.13, but this may be understood due to the lower onset frequency and thus gap magnitude [20] (as marked by the arrows) and the reduced condensate density [19,21]. Our infrared data thus are hardly compatible with the point of view that the SC volume fraction is more than an order of magnitude smaller at x=0.13 than at x=0.15 and x=0.18 where a nearly 100% SC volume fraction is established from µSR [19]. Nevertheless, they do not enable us to accurately determine the SC volume fraction in these underdoped samples. Accordingly, it remains an open question whether the same or different electronic states contribute to the magnetism and to superconductivity and whether these are spatially homogeneous or phase separated on a microscopic or possibly even on a macroscopic scale, if it is limited to a very small volume fraction.

Irrespective of this open question, our data also provide evidence for a mutual interaction between these two kinds of orders. This can be inferred from the doping dependence of the fitting parameters of the ZF-µSR data in Fig. 2(a)-2(c) which show that the onset of superconductivity is accompanied by pronounced changes in the magnetic state. For example, the local magnetic field which is a measure of the magnetic order parameter, after a first drop between x=0 and 0.05, remains almost constant between x=0.05 and 0.1 whereas it exhibits a fairly steep decrease at x>0.1 and vanishes near x=0.15. The related sharp maximum in the relaxation rate around x=0.12-0.13 and the decrease of the static magnetic fraction at x>0.10 are furthermore indicative of a fairly sudden loss of magnetic order combined with the onset of slow dynamical fluctuations. We have already previously shown that these slow magnetic fluctuations persist to the highest achievable doping levels [19]. Superconductivity thus seems to appear as soon as the long-range magnetic order is lost and reaches its maximal value in the presence of slow magnetic fluctuations.

Figure 4 summarises our data in terms of a phase diagram of the superconductivity and magnetism. Also shown are recently reported data of the transition from a tetragonal to orthorhombic structure at low temperature [24], which track the transition temperature of the static magnetism relatively well. This observation agrees with previous reports that the structural transition is a prerequisite for the magnetic one [8]. It shows that the nesting condition of the Fermi-surface which stabilizes the SDW state depends rather sensitively on structural details, such as the Fe-As bond angle. Structural differences may also account for the remarkable differences in the phase diagram with respect to $LaFeAsO_{1-x}F_x$ where a recent μSR study finds that magnetic correlations are limited to a much narrower doping regime and rather abruptly terminate just before SC occurs at x≥0.05 [25]. A similar magnetic phase diagram has been reported in a recent neutron scattering study on $CeFeAsO_{1-x}F_x$ [26] where it was argued that static magnetism and SC also do not coexist. Nevertheless, we note that the structural data yield a similar doping dependence of the tetragonal to orthorhombic transition as in the Sm compound [24, 26]. Accordingly, we expect that future studies with techniques such as μSR, which are very sensitive to weak and strongly disordered magnetism, may reveal a similar coexistence of strongly disordered magnetism and SC as in $SmFeAsO_{1-x}F_x$. These substantial differences in the magnetic and SC phase diagrams of the La and Sm compounds suggest that the magnetic correlations may play an important role in the HTSC pairing mechanism. Notably, in the Sm compound where static magnetism coexists with SC and slow magnetic fluctuations persist to the highest doping levels [19], the maximal $T_c$ is almost double that of the La compound, where static magnetism disappears abruptly at the onset of SC and spin fluctuations are hardly detectable in the SC samples by μSR [25, 27, 28]. While the static magnetic correlations likely do not contribute to the SC pairing interaction, it appears that the slow AF spin fluctuations which emerge in the vicinity of an AF or SDW state may be very beneficial. It is indeed well established that AF spin fluctuations can induce or at least significantly enhance the SC pairing, giving rise to HTSC with an unconventional order parameter [28].

In this context, we also note the similarities with the phase diagram of the cuprate HTSC, where a comparable coexistence of magnetic correlations and superconductivity has been established [14 - 16]. Indeed, strongly disordered static magnetism also coexists with superconductivity in the so-called strongly underdoped regime. Furthermore, in the cuprates the onset of superconductivity gives rise to a suppression of static magnetism and the emergence of slow magnetic fluctuations that persist to higher doping levels, similar to that presented here for $SmFeAsO_{1-x}F_x$. For the cuprate HTSC these cannot be followed up with the μSR technique all the way to optimal doping (due to its limited time window), but they are readily seen for example in inelastic neutron scattering [29].

In summary, we have provided evidence that static magnetism in $SmFeAsO_{1-x}F_x$ survives to surprisingly high doping levels. Strongly disordered but static magnetism and SC both exist in the range of $0.1 \leq x \leq 0.13$ and prominent low energy spin fluctuations are observed up to the highest achievable doping levels where $T_c$ is maximal. The comparison with similar structural studies suggests that the stability of the SDW state critically depends on the structural details that presumably modify the nesting condition at the Fermi-surface or the frustration of competing magnetic interactions. While many of these structural and electronic details in $SmFeAsO_{1-x}F_x$ are very different from those of the cuprate HTSC, we find that the magnetic and superconducting phase diagrams as a function of doping are surprisingly similar. Our observations call for a critical examination of the widely accepted point of view that the HTSC pairing mechanism is based on strong electronic correlations in the vicinity of a Mott-Hubbard-type metal to insulator transition. The comparison with the present SmFeAsO$_{1-x}$F$_x$ superconductors rather points towards an important role of low-energy spin fluctuations that emerge upon doping away from an AF or SDW state.

## Methods

### Sample preparation and characterisation

Polycrystalline samples with nominal composition SmFeAsO$_{1-x}$F$_x$ with x=0.00, 0.05, 0.10, 0.12, 0.13, 0.15, 0.18 and 0.30 were synthesized by conventional solid state reaction methods as described in Ref [2,11]. Standard powder x-ray diffraction patterns were measured where all peaks could be indexed to the tetragonal ZrCuSiAs-type structure. DC resistivity and magnetisation measurements were made to determine the midpoint (10% to 90% width) of the resistive and diamagnetic transitions with T$_c$=10(7), 17(8), 25(8), 38(4), 45(3) and 45(4) K for x=0.1, 0.12, 0.13, 0.15, 0.18, and 0.3 respectively.

### Muon spin rotation

The 100% spin polarized muon spin rotation/relaxation (µSR) experiments were performed using the GPS and LTF spectrometers at the Paul Scherer Institut (PSI) Switzerland for the low doping samples with 0 ≤ x ≤ 0.15 and using the EMU, ARGUS and MUSR spectrometers at the ISIS Facility UK for the samples with 0.15 ≤ x ≤ 0.30. The µSR technique is especially suited for the study of magnetic and superconducting (SC) materials since it allows one to study the local magnetic field distribution on a microscopic scale and thus to directly access the volume fractions of the corresponding phases [28]. For example in the case of the strongly underdoped cuprate HTSC this technique has been very successfully applied to reveal the coexistence of weak and strongly disordered magnetism and superconductivity [14,15,29,30]. Details of this technique are outlined in the supporting online material.

## Acknowledgements

This work is supported by the Schweizer Nationalfonds (SNF) by grant 200020-119784 and the NCCR program MANEP, the Deutsche Forschungsgemeinschaft (DFG) by grant BE2684/1-3 in FOR538 and the UK EPSRC. We acknowledge helpful discussions with D. Baeriswyl, A.T. Boothroyd and M. Siegrist.

## Additional Information

Supplementary Information accompanies this paper on www.nature.com/naturematerials. Correspondence and requests for materials should be addressed to C.B or A.J.D.

## Author Contributions

A.J.D, Ch. N., F. L. P., S. J. B., T. L., I. W., C. B. 1 and C. B 2 performed the muon experiments. A.J.D, Ch. N., F. L. P., S. J. B. and C. B 2 analysed and interpreted the, A. J. D., P. J. B., S. J. B., V. K. M, A. D., M. R., W. K., and C. B. 2 were responsible for the characterisation measurements, R. H. L., G. W and X. H. C prepared the samples.

Figure 1: **Temperature- and doping dependence of the zero-field μSR spectra of SmFeAsO$_{1-x}$F$_x$.** Time dependent spectra of the muon-spin-polarisation, P(t)/P(0), at representative temperatures below and above the magnetic ordering temperature, T$_{mag}$, for polycrystalline SmFeAsO$_{1-x}$F$_x$ samples with (a) x=0 and T$_{mag}$≈135 K, (b) x=0.05 and T$_{mag}$≈80 K, (c) x=0.1, T$_{mag}$≈60 K and T$_c$=10(7) K, (d) x=0.12, T$_{mag}$≈35 K and T$_c$=17(8) K, and (e) x=0.13, T$_{mag}$≈30 K and T$_c$=25(8) K. Symbols show experimental data and solid lines fits with Equation 1 for which the obtained parameters are shown in Figs. 2(a)-2(c).

Figure 2: **Evolution of the magnetic signal of the μSR measurements on SmFeAsO$_{1-x}$F$_x$ as a function of doping and temperature** (a) Doping dependence of the local magnetic field, B$_\mu$, and the relaxation rate, $\lambda^{ZF}$, as obtained from fitting the ZF-μSR data. The red shaded area marks the SC region. (b) Doping dependence of the amplitudes f$_1$ and f$_2$ in the ZF-μSR spectra together with the superconducting transition temperatures, T$_c$, as obtained from resistivity and magnetisation measurements. Inset: magnetic volume fraction f$_1$+f$_2$ obtained from the low temperature weak transverse field μSR measurements (c) Temperature dependence of the local field at the muon site, B$_\mu$, as deduced from the precession frequency, ν$_\mu$, in the ZF-μSR spectra with ν$_\mu$=(γ$_\mu$B$_\mu$/2π) and γ$_\mu$=2 π*135.3 MHz/T the gyromagnetic ratio of the muon. Arrows indicate our estimates of the magnetic transition temperature, T$_{mag}$.

Figure 3: **Evidence for bulk superconductivity due to a gap-like suppression of the infrared optical conductivity, and magnetisation and resistivity data supporting the presence of superconductivity.** (a) Difference spectra of the far-infrared optical conductivity between the normal and the superconducting states for SmFeAsO$_{1-x}$F$_x$ with x=0.13, 0.15 and 0.18 and T$_c$=25(8) K, T$_c$=38(4) K, and T$_c$=45(3) K, respectively. The gap-like suppression is a clear signature of bulk superconductivity. The onset frequency as marked by the arrows is roughly proportional to the SC gap energy. (b) Magnetisation data for our data, where a sizeable diamagnetic shift is observed for all of the superconducting samples. (c) Resistivity data, scaled to the resistivity at 280K, for all of our samples.

Figure 4: **Phase diagram of the magnetic and superconducting properties of SmFeAsO$_{1-x}$F$_x$.** Evolution of the main magnetic transition temperature, $T_{mag}$ (blue squares), the Sm ordering temperature $T_{Sm}$ (grey triangles [18]), the superconducting transition temperature $T_c$ (red circles) and the structural transition $T_s$ (green triangles [24]), as a function of the F substitution and thus electron doping. There is a clear region of coexistence between x=0.10 and x=0.15 and $T_c$ reaches its maximal value just as the main static magnetism phase disappears.

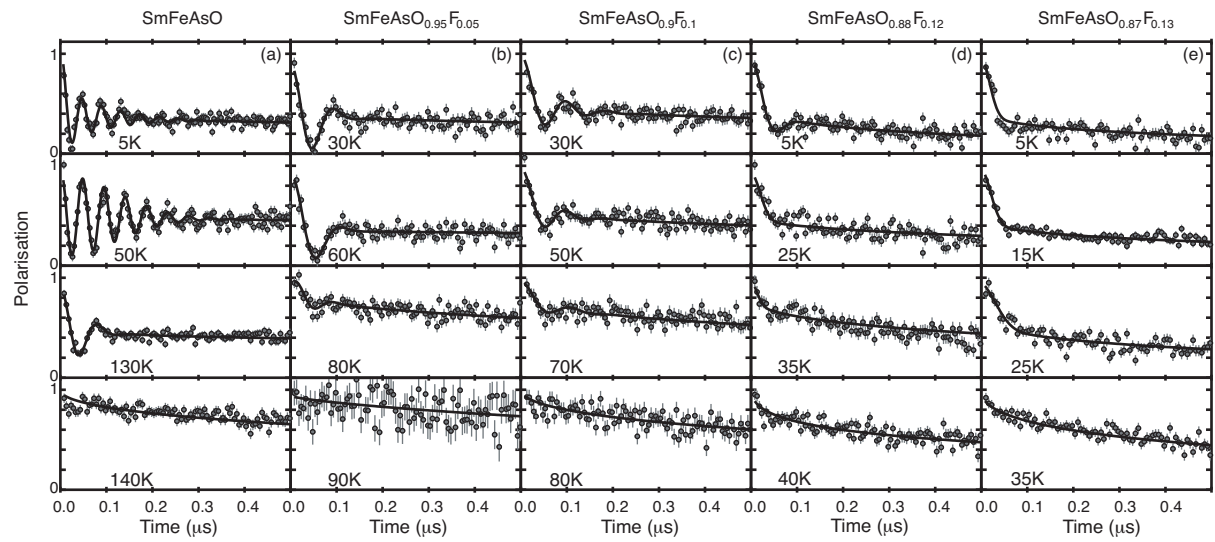

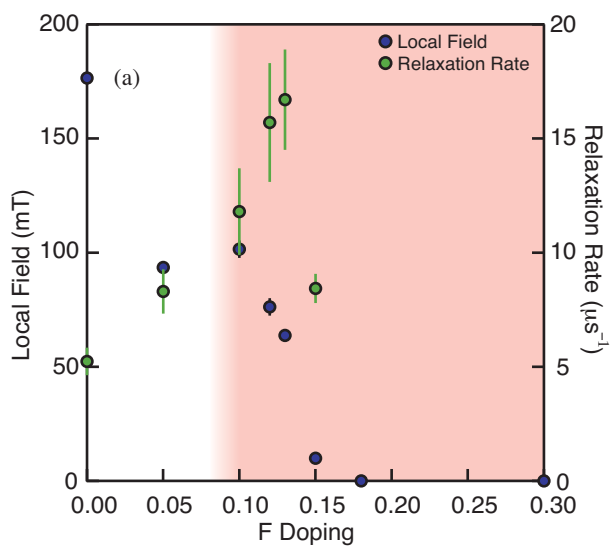
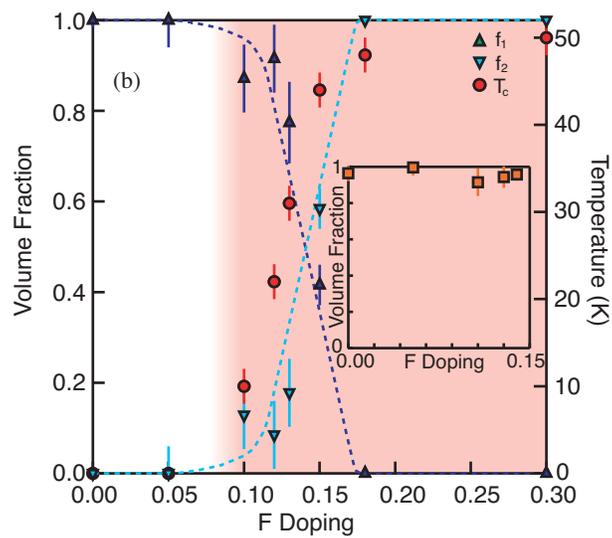
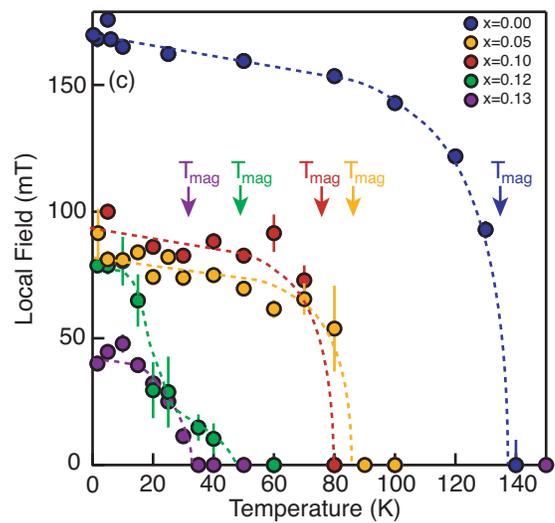

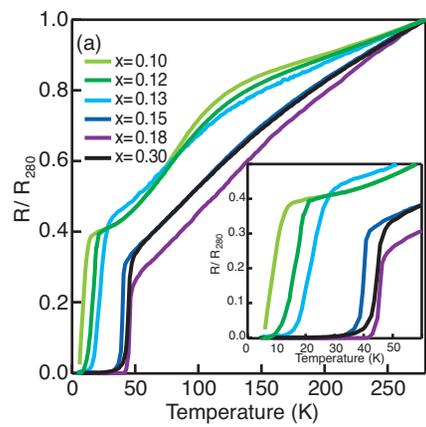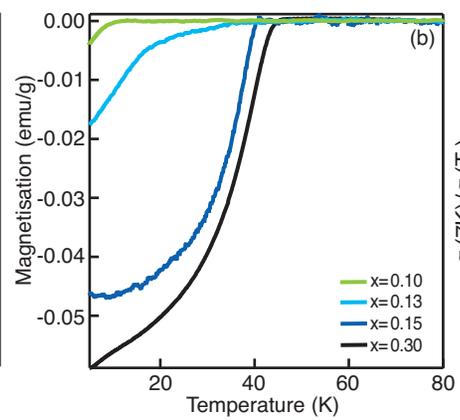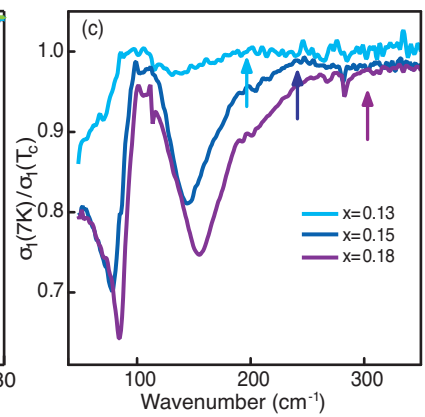

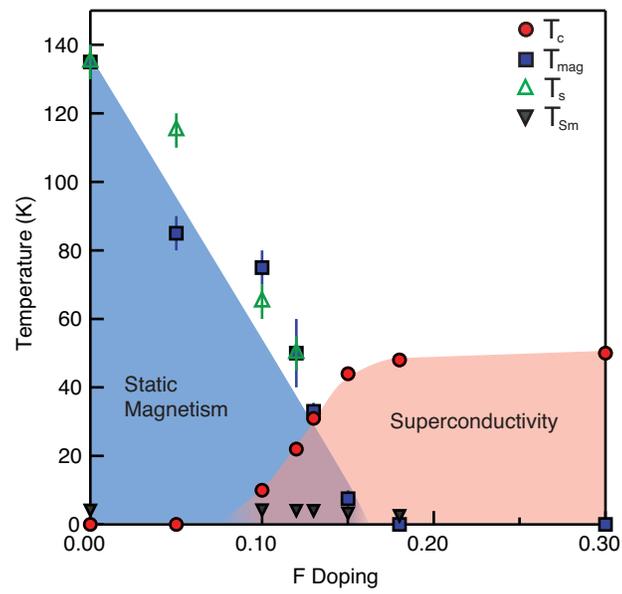

**Section 1: Muon Spin Rotation**

Muons are elementary particles belonging to a family of particles called leptons that includes the more familiar electron. One can think of the positive muon as a light proton with a mass of one ninth of the proton mass and magnetic moment approximately 3.2 times larger than that of the proton. The availability of a spin polarized 4.1 MeV positive-muon beam has opened up the possibility to use muons as an extremely sensitive and versatile magnetic microprobe. Surface muons have an energy of 4.1 MeV and they therefore have a stopping range in a solid that varies from 0.1 to 1 mm.

Positive muons decay to a positron, muon antineutrino and electron neutrino. The angular emission of positrons is well characterised, with the emission direction being correlated with the muon's spin at time of decay. Thus, by measuring the direction and the timing of a statistically significant number of decay positrons it is possible to follow directly the evolution of the muon's spin ensemble as a function of time after implantation. This allows a wealth of information to be gained about the host material in which the 100% spin polarised muons are implanted and come to rest. They can act as passive local magnetic microprobes, for example directly measuring magnetic field distribution at the implanted site with very high sensitivity (less than 0.1 mT). Being able to follow the evolution of the spin with time means that the magnetic field experienced by the muon can be determined through the measurement of the Larmor precession of the muon spin. In a magnetic field the spin will precess about the field direction with a frequency $\omega_\mu$ proportional to the field $B$

$$\omega_\mu = \gamma_\mu B \tag{2.1}$$

where $\gamma_\mu/2\pi = 135.5$ MHz T$^{-1}$ is the gyromagnetic ratio for the muon.

The spin rotation can be observed using two (or more) positron counters, *a* and *b*, mounted on the opposite sides of the sample. The number of positrons detected by each counter as a function of time ($H^a(t)$ and $H^b(t)$) reflects the time dependence of the muon spin polarisation along the axis of observation defined by the two detectors:

$$H^a(t) = N_0^a \cdot \exp(-t^a/\tau_\mu) \cdot [1 + A^a(t)] + C^a \quad (2.2)$$

$$H^b(t) = N_0^b \cdot \exp(-t^b/\tau_\mu) \cdot [1 + A^b(t)] + C^b \quad (2.3)$$

where $N_0^{a,b}$ is the initial counts at zero time, $\tau_\mu \sim 2.2\mu s$ is the muon's lifetime, $C^{a,b}$ is the time independent background and the asymmetry $A^{a,b}(t)$. The asymmetry contains all of the information about the time evolution of the muon's spin polarisation.

**Section 2: Low Temperature Ordering of Sm Moments**

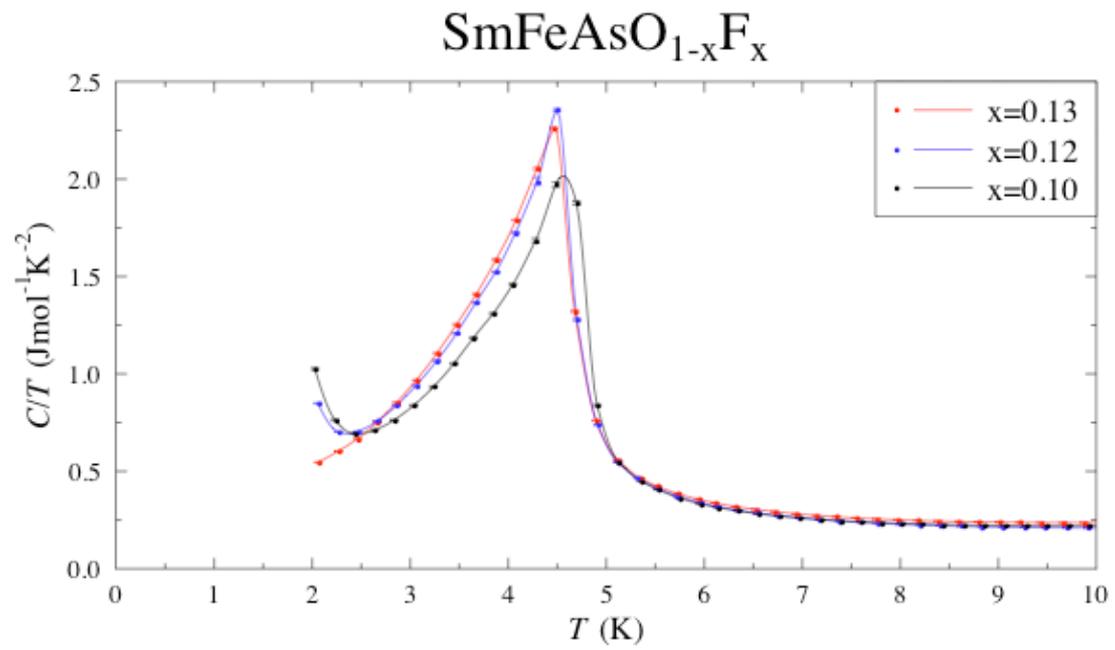

*Figure 1: Heat capacity measurements showing the Sm moments ordering at low temperatures for three different dopings.*

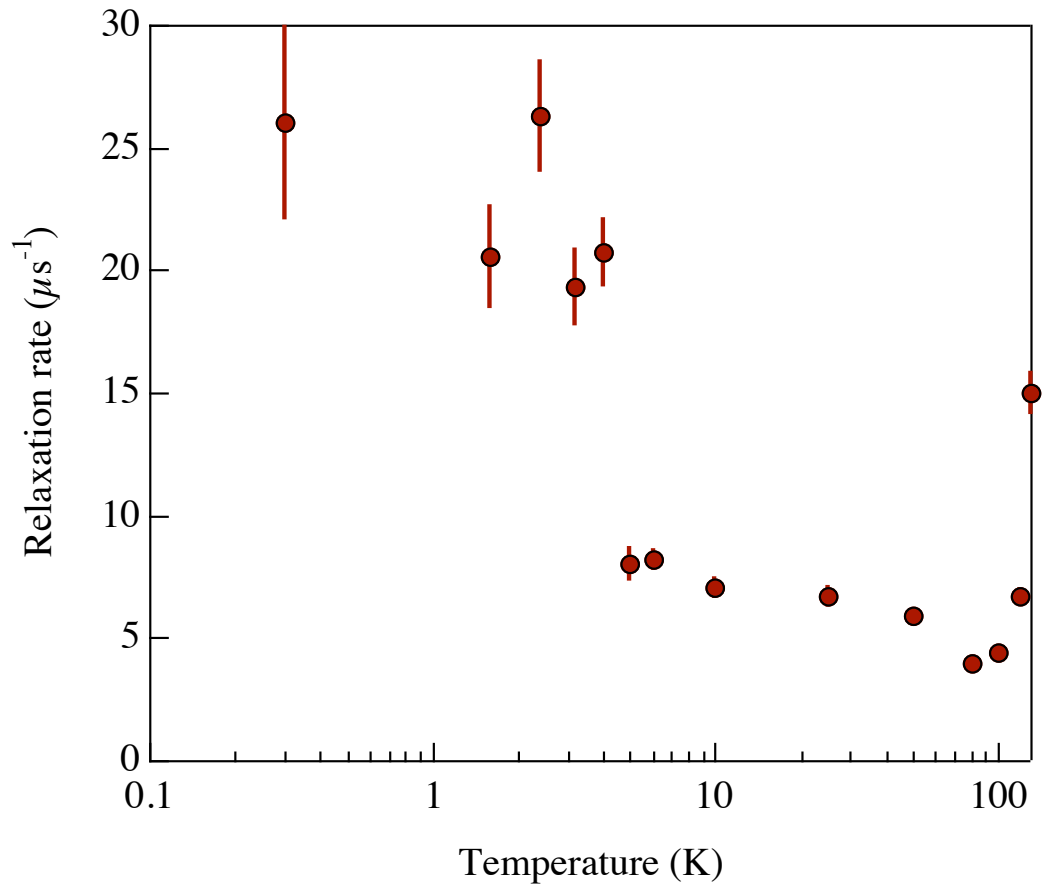

*Figure 2: The Gaussian relaxation rate of the 23MHz oscillation (in the undoped parent compound), where a clear and dramatic increase is observed at T < 5 K, corresponding to the Sm moments ordering.*

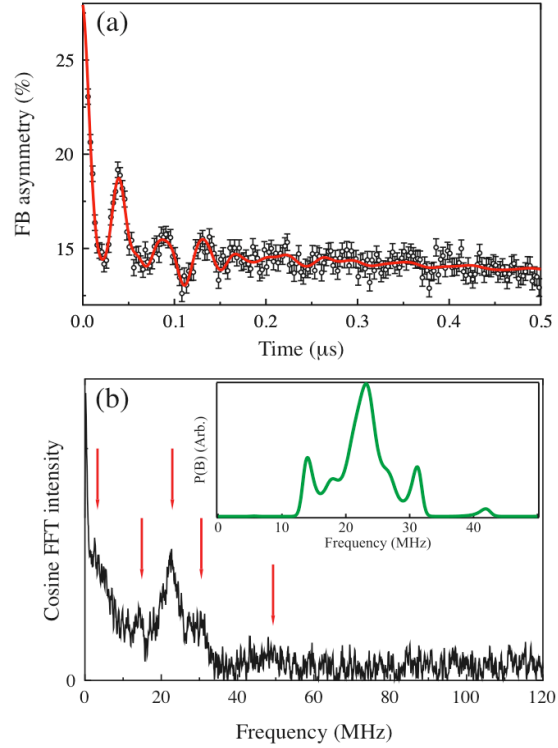

*Figure 3: ZF-mSR spectra and analysis for the undoped parent compound at 0.3K. (a) A fit (red line) to the time domain data (black points) using a five component Gaussian relaxed oscillation. (b) The Cosine transform of the data shown in (a), where five peaks are clearly observed. The red arrows mark the frequencies obtained from the time domain fits. The inset shows a maximum entropy transform of the data shown in (a). We note that the maximum entropy algorithm has difficulty resolving low frequencies, and as a consequence only the four highest frequencies are observable. The additional frequencies observed at 0.3K are not present in the higher temperature data, which only show the 23 MHz signal (Fig. 1a of the main paper), suggesting that they are related to the ordering of the Sm moments at low temperatures.*